\documentclass[12pt,
aps,
floatfix,
pra,reprint,
 superscriptaddress,
 numerical,
longbibliography
]{revtex4-1}
\usepackage{dcolumn}
\usepackage{bm}
\usepackage%
{graphicx}
\usepackage{xfrac}
\usepackage{extdash}
\usepackage{epsfig}
\usepackage{isomath}
\usepackage{textcomp}
\usepackage{mathrsfs}
\usepackage{bm}
\usepackage[english]{babel}
\usepackage{color,soul}
\usepackage{amsfonts}
\usepackage[columnwise]{lineno}
\usepackage{gensymb}
\usepackage{bm}
\usepackage{amsmath}

\usepackage{tikz}
\usepackage{pgfmath}
\usepackage{pgfplots}
\pgfplotsset{compat=1.18}
\usetikzlibrary{intersections,backgrounds}

\usepackage{times}
\usepackage[normalem]{ulem}

\DeclareFontFamily{U}{euc}{}
\DeclareFontShape{U}{euc}{m}{n}{<-6>eurm5<6-8>eurm7<8->eurm10}{}%
\DeclareSymbolFont{AMSc}{U}{euc}{m}{n} 
\DeclareMathSymbol{\umu}{\mathord}{AMSc}{"16}

\newcommand{\ensuretext}[1]{\ensuremath{\text{#1}}}

\newcommand{\sign}{\ensuremath{\mathrm{sign}}}

\newcommand{\unit}[1]{\ensuretext{\textrm{\,}}\ensuremath{\mathrm{#1}}}

\newcommand{\Mum}{\ensuremath{\umu}\ensuremath{\mathrm{m}}}
\newcommand{\mum}{\textrm{\,\ensuremath{\mathrm{\Mum}}}}

\usepackage{xurl}
\usepackage[hidelinks]{hyperref}
\begin{document}

\title{Microscopy X-ray Imaging enriched with Small Angle X-ray Scattering for few nanometre resolution reveals shock waves and compression in intense short pulse laser irradiation of solids} 

\author{Thomas Kluge$^{1,}$}
\email{t.kluge@hzdr.de} 
\affiliation{Helmholtz-Zentrum Dresden-Rossendorf, Bautzner Landstra\ss e 400, 01328, Dresden, Germany}

\author{Arthur Hirsch-Passicos}
\affiliation{Helmholtz-Zentrum Dresden-Rossendorf, Bautzner Landstra\ss e 400, 01328, Dresden, Germany}

\author{Jannis Schulz}
\affiliation{Helmholtz-Zentrum Dresden-Rossendorf, Bautzner Landstra\ss e 400, 01328, Dresden, Germany}
\affiliation{Technical University Dresden, 01069 Dresden, Germany}

\author{Mungo Frost}
\affiliation{SLAC National Accelerator Laboratory, 2575 Sand Hill Rd, Menlo Park, CA 94025, USA}

\author{Eric Galtier}
\affiliation{SLAC National Accelerator Laboratory, 2575 Sand Hill Rd, Menlo Park, CA 94025, USA}

\author{Maxence Gauthier}
\affiliation{SLAC National Accelerator Laboratory, 2575 Sand Hill Rd, Menlo Park, CA 94025, USA}

\author{J\"org Grenzer}
\affiliation{Helmholtz-Zentrum Dresden-Rossendorf, Bautzner Landstra\ss e 400, 01328, Dresden, Germany}

\author{Christian Gutt}
\affiliation{Universit\"at Siegen, Walter-Flex Stra\ss e 3, 57072 Siegen, Germany}

\author{Lingen Huang}
\affiliation{Helmholtz-Zentrum Dresden-Rossendorf, Bautzner Landstra\ss e 400, 01328, Dresden, Germany}

\author{Uwe H\"ubner}
\affiliation{Leibniz Institute of Photonic Technology, Albert-Einstein-Stra\ss e 9, 07745 Jena, Germany}

\author{Megan Ikeya}
\affiliation{SLAC National Accelerator Laboratory, 2575 Sand Hill Rd, Menlo Park, CA 94025, USA}

\author{Hae Ja Lee}
\affiliation{SLAC National Accelerator Laboratory, 2575 Sand Hill Rd, Menlo Park, CA 94025, USA}

\author{Dimitri Khaghani}
\affiliation{SLAC National Accelerator Laboratory, 2575 Sand Hill Rd, Menlo Park, CA 94025, USA}

\author{Willow Moon Martin}
\affiliation{SLAC National Accelerator Laboratory, 2575 Sand Hill Rd, Menlo Park, CA 94025, USA}

\author{Brian Edward Marré}
\affiliation{Helmholtz-Zentrum Dresden-Rossendorf, Bautzner Landstra\ss e 400, 01328, Dresden, Germany}
\affiliation{Technical University Dresden, 01069 Dresden, Germany}

\author{Motoaki Nakatsutsumi}
\affiliation{European XFEL, Holzkoppel 4, 22869 Schenefeld, Germany}

\author{Paweł Ordyna}
\affiliation{Helmholtz-Zentrum Dresden-Rossendorf, Bautzner Landstra\ss e 400, 01328, Dresden, Germany}
\affiliation{Technical University Dresden, 01069 Dresden, Germany}

\author{Franziska-Luise Paschke-Br\"uhl}
\affiliation{Helmholtz-Zentrum Dresden-Rossendorf, Bautzner Landstra\ss e 400, 01328, Dresden, Germany}

\author{Alexander Pelka}
\affiliation{Helmholtz-Zentrum Dresden-Rossendorf, Bautzner Landstra\ss e 400, 01328, Dresden, Germany}

\author{Lisa Randolph}
\affiliation{Universit\"at Siegen, Walter-Flex Stra\ss e 3, 57072 Siegen, Germany}

\author{Hans-Peter Schlenvoigt}
\affiliation{Helmholtz-Zentrum Dresden-Rossendorf, Bautzner Landstra\ss e 400, 01328, Dresden, Germany}

\author{Christopher Schoenwaelder}
\affiliation{SLAC National Accelerator Laboratory, 2575 Sand Hill Rd, Menlo Park, CA 94025, USA}

\author{Michal \v{S}m\'{i}d}
\affiliation{Helmholtz-Zentrum Dresden-Rossendorf, Bautzner Landstra\ss e 400, 01328, Dresden, Germany}

\author{Long Yang}
\affiliation{Helmholtz-Zentrum Dresden-Rossendorf, Bautzner Landstra\ss e 400, 01328, Dresden, Germany}

\author{Ulrich Schramm}
\author{Thomas E. Cowan}
\affiliation{Helmholtz-Zentrum Dresden-Rossendorf, Bautzner Landstra\ss e 400, 01328, Dresden, Germany}
\affiliation{Technical University Dresden, 01069 Dresden, Germany}

\date{\today}

\begin{abstract}
Understanding how laser pulses compress solids into high-energy-density states requires diagnostics that simultaneously resolve macroscopic geometry and nanometre-scale structure. Here we present a combine X-ray imaging (XRM) and small-angle X-ray scattering (SAXS) approach that bridges this diagnostic gap. Using the Matter in Extreme Conditions end station at LCLS, we irradiated $25~\mu\mathrm{m}$ copper wires with $45~\mathrm{fs}$, $0.9~\mathrm{J}$, $800~\mathrm{nm}$ pulses at $3.5\times10^{19}~\mathrm{W/cm^2}$ while probing with $8.2~\mathrm{keV}$ XFEL pulses. XRM visualizes the evolution of ablation, compression, and inward-propagating fronts with $\sim200~\mathrm{nm}$ resolution, while SAXS quantifies their nanometre-scale sharpness via the time-resolved evolution of scattering streaks. The joint analysis reveals that an initially smooth compression steepens into a nanometre-sharp shock front after $t_{\mathrm{sh}}\approx(18\pm3)~\mathrm{ps}$, consistent with an analytical steepening model and FLASH hydrodynamic simulations. The front reaches a velocity of $c_{\mathrm{sh}}\approx25~\mathrm{km/s}$ and a lateral width of several tens of microns, demonstrating for the first time direct observation of shock formation and decay at solid density with few-nanometre precision. This integrated XRM–SAXS method establishes a quantitative, multi-scale diagnostic of laser-driven shocks in dense plasmas relevant to inertial confinement fusion, warm dense matter, and planetary physics.

\end{abstract}
\maketitle 

\section*{Introduction}

The controlled compression of matter to extreme pressures and densities is a central objective in high-energy-density (HED) and inertial confinement fusion (ICF) research. Conventional approaches rely on long-pulse, high-energy lasers to compress spherical fuel capsules either indirectly, via X-rays generated in a hohlraum, or directly by irradiating the capsule surface. While these schemes have achieved remarkable success---notably ignition at the National Ignition Facility---they require large, costly facilities, which limits experimental access and constrains the development of new target concepts and diagnostics.

Alternative routes based on short-pulse, high-intensity lasers have recently gained attention as compact drivers for transient, extreme compression~\cite{Tabak1994,Roth2001,Ruhl2022,Garcia2024}. In the latter case, an ultra-intense pulse irradiates a solid target, typically a thin wire or planar foil, generating energetic electrons that drive return currents in the bulk. The associated magnetic fields and ablation counterpressures launch inward-moving compression waves that can steepen into shocks. For cylindrical wires, upon convergence on the target axis these shocks can produce Gbar-level pressures and strongly heated, highly ionized states of matter---conditions directly relevant for ICF physics, planetary interiors, and laboratory astrophysics.

The ability to characterise these laser-driven shocks on multiple scales is critical. X-ray microscopy (XRM) in direct imaging geometry\cite{Galtier2025} has enabled visualisation of plasma density evolution including blast waves, hole-boring, wire implosions, and filamentation with spatial resolutions down to a few hundred nanometres~\cite{Garcia2024,Schoenwaelder2025}. However, even this resolution is insufficient to resolve the nanometre-scale sharpness of shock fronts themselves. Small-angle X-ray scattering (SAXS), in contrast, can provide sensitivity to structural variations at the nanometre scale and has recently been applied to diagnose phase interfaces and density gradients in laser-compressed solids~\cite{Kluge2017,kluge2018,Gaus2021,Kluge2023}. 
Yet, previous SAXS experiments suffered from two main limitations: the scattering signal was often too weak for quantitative analysis, and the lack of complementary real-space imaging left the interpretation of the scattering features ambiguous.

\begin{figure}
\centering
  \includegraphics[width=\linewidth]{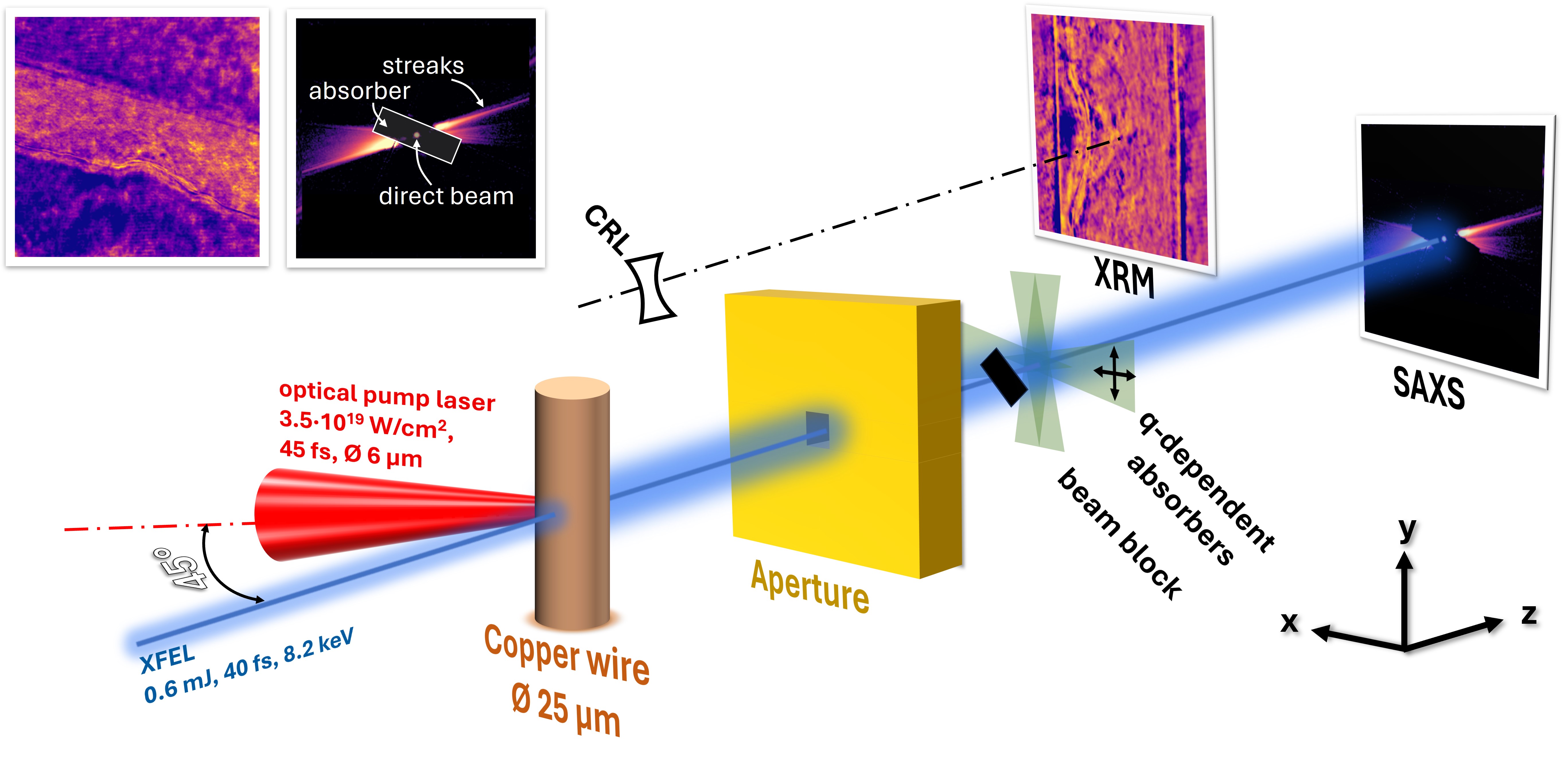}
\caption{Experimental setup. The MEC HI laser (red) is focused onto the Copper wire target under $45^\circ$ in p-polarisation, the XFEL (blue) is probing the plasma density. The detector records the Small-angle X-ray scattering (SAXS) image normalised at small $q$ values by the retractable absorber system. 
The diagnostics are on motorised stages so the absorbers, beam block and SAXS detector can be easily exchanged for the CRL stack and imaging detector. The figure is not to scale. The insets show raw images for XRM and SAXS of HI-laser driven wires after $100\unit{ps}$. }
  \label{fig:setup}
\end{figure}
In this work, we overcome these limitations by combining XRM and SAXS in a joint, spatially correlated diagnostic setup. XRM provides real-space constraints on the macroscopic morphology and evolution of the compression fronts, while SAXS delivers quantitative information on their nanometre-scale sharpness and evolution in time. This integrated approach enables simultaneous access to global geometry and local structure, bridging the gap between hydrodynamic and microscopic descriptions of laser-driven compression.

\begin{figure*}
\centering
\includegraphics[width=\linewidth]{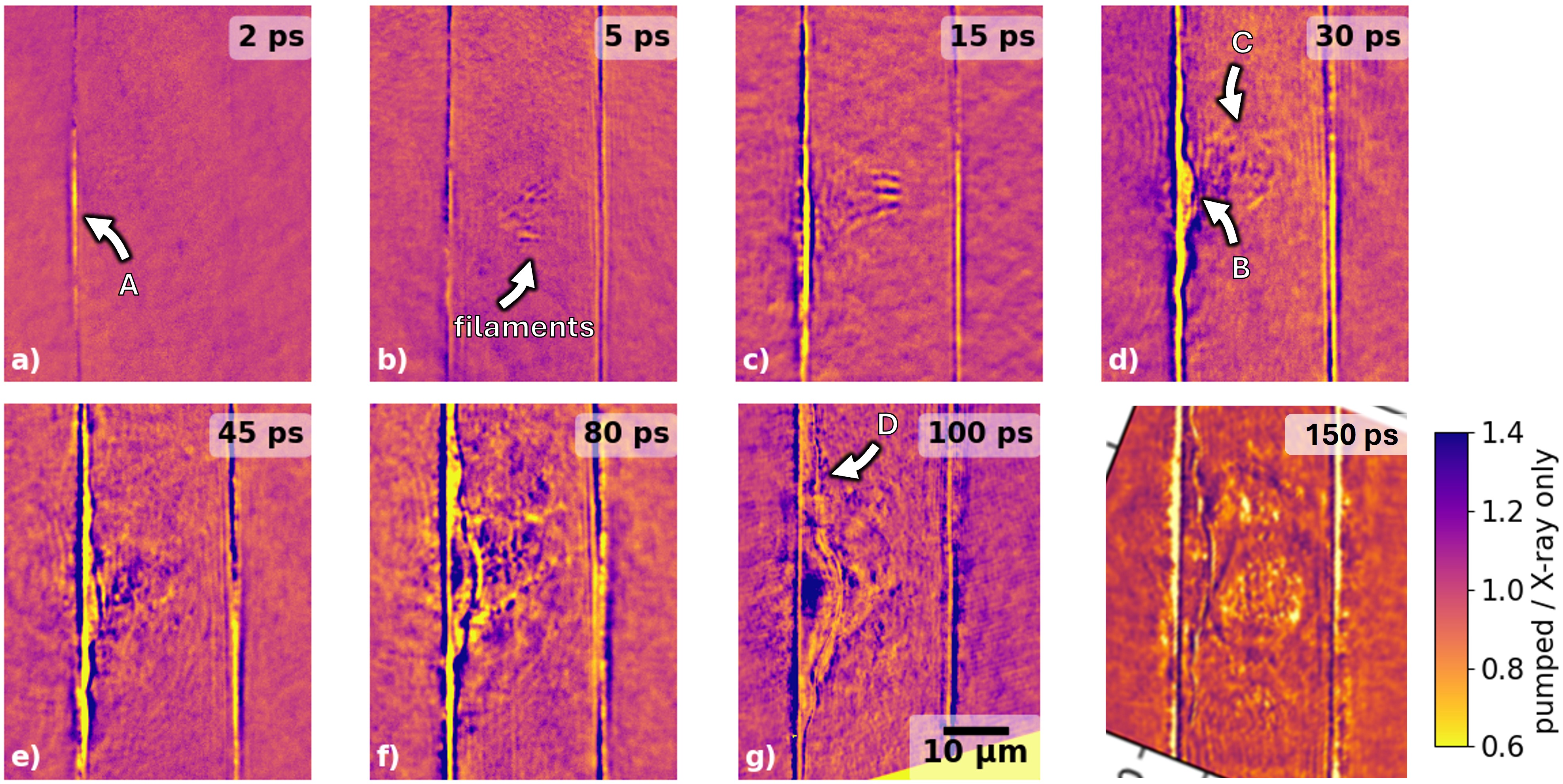}
\caption{X-ray microscopy  results.
Sequence of representative XRM images showing the evolution of the Cu wire after laser irradiation (shown is the ratio of the pumped wire image over the cold image).  
Distinct features (A–D) mark ablation, surface compression, hole-boring-like fronts, and inward-propagating compression fronts, respectively. See main text for details. }
\label{fig:MXI}
\end{figure*}

The experiment was conducted at the Matter in Extreme Conditions (MEC) end station of the Linac Coherent Light Source (LCLS) X-ray free-electron laser (XFEL) at SLAC National Accelerator Laboratory, see Fig.~\ref{fig:setup}. 
We irradiated copper wires of $25\unit{\mum}$ diameter with $45\unit{fs}$, $0.9~\unit{J}$, $800\unit{nm}$ pulses from the MEC short-pulse high-intensity (HI) laser, focused to a spot of  $6\unit{\mum}$ full width at half maximum (FWHM). 
This corresponds to a peak intensity of $3.5\times10^{19}\unit{W/cm^2}$ and a normalised vector potential $a_0=4$. 
The plasma dynamics were probed by $40\unit{fs}$ XFEL pulses ($E_\gamma=8.2~\mathrm{keV}$, $\sim0.6\unit{mJ}$ per pulse). 
Both beams were focused to the wire cylinder surface, with $45^\circ$ angle encompassed between them. 
The SAXS geometry was optimised for high photon fluence and dynamic range, employing a $15\unit{\mum}$ XFEL spot size on target that was centered to the HI laser side of the target. 
After a $5\unit{m}$ propagation distance the X-ray signal was recorded with an ePix10k-540k detector behind a beam stop and adjustable attenuation system to control the low-$q$ signal. 
This allowed full use of the XFEL intensity without detector saturation and coverage of a large $q$-range down to small $q$ values close to the XFEL beam waist, a major improvement over earlier experiments that required severe flux reduction~\cite{Kluge2023}.

For XRM, we used the MEC X-ray imager (MXI) platform\cite{Galtier2025}. 
A larger XFEL spot of approx. $150\unit{\mum}$ was used to illuminate a broader target region while protecting the downstream Beryllium compound refractive lens stack (CRL) used to image the target to an Andor Neo 5.5 sCMOS scintillator camera. 
The configuration followed the high-resolution setup described in~\cite{Schoenwaelder2025}, achieving a spatial resolution of $200\unit{nm}$, as verified with a Siemens star test pattern. 
Note that due to the setup the laser interaction region appears inside the target in the XFEL projection. 
The change from SAXS to MXI setup could be facilitated in less than $20\unit{min}$. 
Together, this dual-diagnostic scheme provides a comprehensive, quantitative probe of laser-driven compression from tens of microns to nanometres---a capability essential for disentangling the formation, evolution, and decay of shock fronts in dense plasmas.

\section*{Results}

\begin{figure*}
\centering
\includegraphics[width=\linewidth]{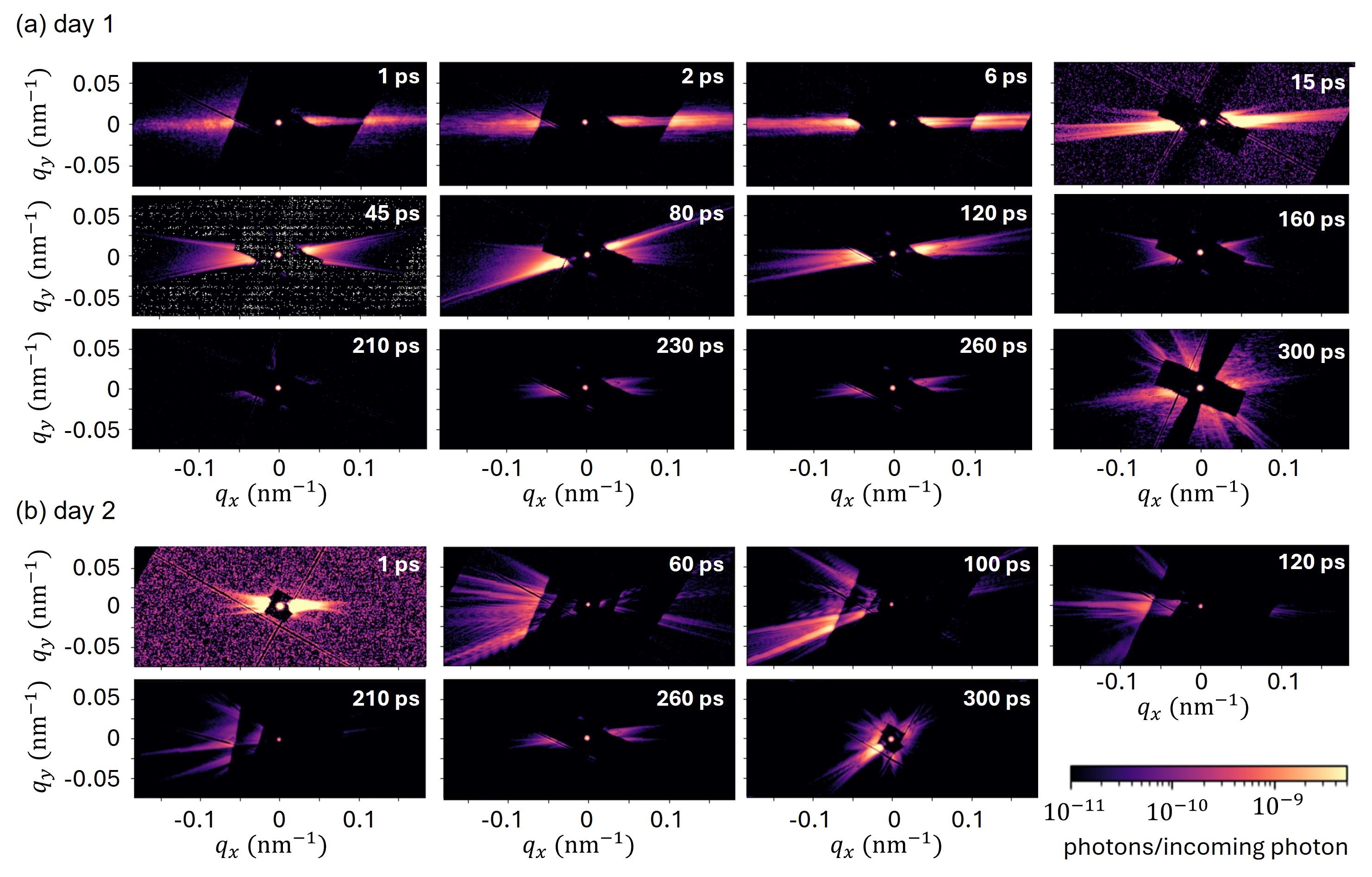}
\caption{Evolution of the SAXS pattern for representative probe delays grouped by the two experiment days (a) day 1 and (b) day 2. 
The tilt, number, and sharpness of streaks increase with time, reflecting the evolving curvature, multiplicity of compression fronts, and shock formation. Note that for each image the X-ray absorber positions are slightly different for an individual optimisation. }
\label{fig:SAXS}
\end{figure*}
\subsection*{X-ray microscopy (XRM)}
Figure~\ref{fig:MXI} summarises representative XRM data obtained for probe delays between $2\unit{ps}$ and $100\unit{ps}$ after the short-pulse laser irradiation. The overall evolution reveals a rapid onset of surface deformation, followed by the emergence of distinct compression and recession fronts. As early as $2\unit{ps}$, a small reduction in transmission appears on the laser-facing side of the wire (hereafter referred to as the \emph{front} surface), extending roughly $\pm$10\unit{\mum} transversely around the laser axis. By $5\unit{ps}$, this feature spans the full transverse field of view and becomes visible also on the opposite (\emph{rear}) side. Simultaneously, a zone of increased transmission develops toward the vacuum interface (region~A in Fig.~\ref{fig:MXI}). Such contrast changes are consistent with surface ablation and recession accompanied by local compression of the adjacent material---a process driven by the combination of return-current heating and magnetic counter-pressure~\cite{Garcia2024}. Bound–bound absorption near the Cu~K$\alpha$ line at 8~keV may further contribute to the apparent opacity increase in the compressed regions~\cite{Kluge2016}.

At later times ($t\gtrsim30\unit{ps}$), the front surface begins to curve inward, forming a quasi-Gaussian profile that deepens progressively into the bulk (region~B). Around $80-100\unit{ps}$, this deformation detaches from the surface, indicating the propagation of a distinct inward-moving compression front. 
This front appears to be best described by two Gaussians: a narrow one similar to the laser focus width and a wider one with several tens of microns width describing the shallow rather straight front at a transverse offset. 
This latter front moves inwards with a velocity of a few tens of km/s (region~D). The front reaches a maximum inclination of approximately $15^{\circ}$ near $80\unit{ps}$ and subsequently relaxes. 

Concurrently, a pronounced, turbulent zone forms around the laser impact point, featuring filamentary modulations with characteristic wavelengths of $\lambda_f \approx 1\unit{\mum}$ (after $5\unit{ps}$) to $1.5\unit{\mum}$ (after $15\unit{ps}$). These modulations coincide spatially with the expected position of the laser–plasma interaction and are consistent with resistive filamentation observed under similar conditions~\cite{Schoenwaelder2025}.
A distinct, hole-boring-like feature
\footnote{Note, that the compression surface is likely a shock front following the hole boring acceleration during the laser irradiation, not the hole-boring process itself (which would be limited only to the laser irradiation phase). For better readability we refer to it with the term hole-boring front.} 
(region~C) appears around the focal projection at $\sim30\unit{ps}$, which expands in diameter over time and evolves into a blast-wave-like structure beyond $100\unit{ps}$. 
The region behind this front shows significant shot-to-shot variability, indicative of hydrodynamic instabilities and mixing between compressed and unperturbed material.

Interpretation of the XRM data requires care due to the 45$^{\circ}$ inclination between the pump and probe beams. 
The laser interacts with the cylindrical surface at normal incidence, so that the apparent “left” and “right” wire surfaces in projection do not correspond directly to the true front and rear surfaces w.r.t. the laser incidence. 
To clarify the correspondence, we constructed an analytic wire-density model and compared its projections with the experimental images. The model parameterised the two fronts by two superimposed Gaussian density perturbations, respectively (one Gaussian for the more localised hole-boring fronts (B,C), and an additional Gaussian for the wide compression front (D)), whose projected density distributions reproduce the observed transmission profiles (see Fig.~\ref{fig:saxs-mxi}(b) for an example at $100\unit{ps}$ delay).

Despite the lower drive intensity compared with the cylindrical-convergence experiment reported in~\cite{Garcia2024}, the same fundamental physical processes seem to appear here. 
In particular, an inward traveling compression front appears at the surface, that extends quite far transversely even several times the laser focal width away from the laser axis, and which is consistent with either a shock front driven by return-current heating or direct heating by out-of-focus laser intensity.
In the present experiment with less pump laser intensity than that used in~\cite{Garcia2024} the propagation ceases after $\sim100~\mathrm{ps}$.  

\subsection*{Small-angle X-ray scattering (SAXS)}

Complementary SAXS data are presented in Fig.~\ref{fig:SAXS} for various probe delays. At $6\unit{ps}$, vertical streaks perpendicular to the target surface begin to split and tilt, indicating curvature of the scattering interface. With increasing delay, the tilt angle grows and multiple streaks appear, particularly in the data from the second experimental day. In contrast to earlier, lower-intensity experiments---which exhibited numerous streaks attributed to small-scale surface ripples~\cite{Kluge2017}---the present data show only a few, well-defined streaks corresponding to larger, more controlled structures.

\begin{figure*}
\centering
\includegraphics[width=\linewidth]{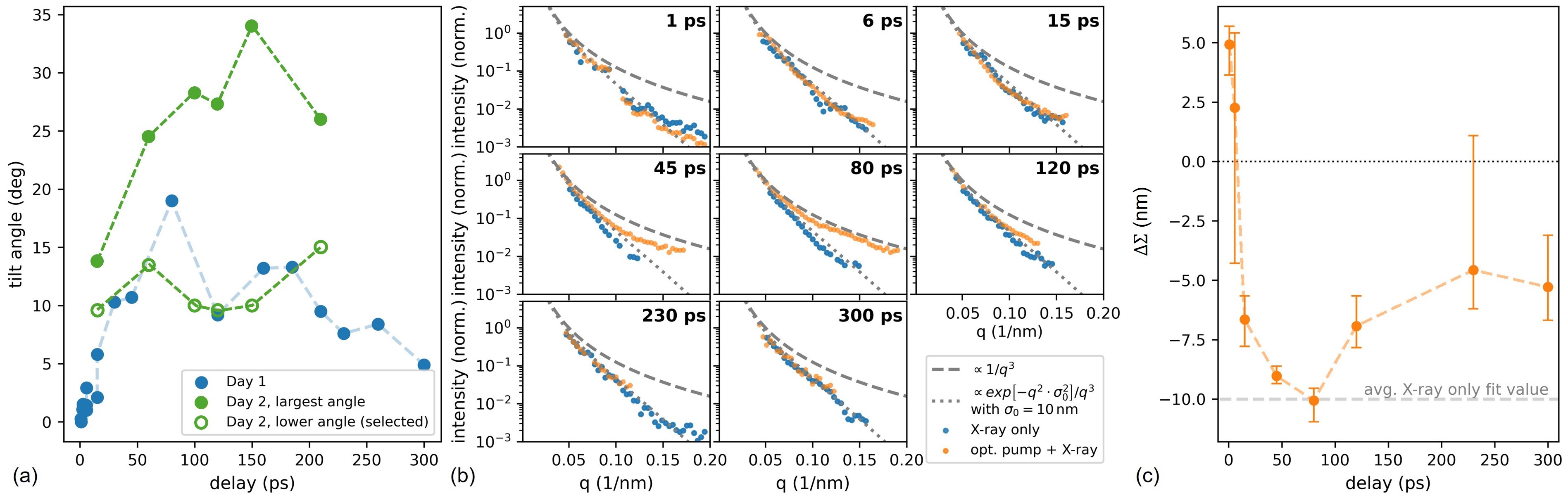}
\caption{(a) Tilt angle of SAXS scattering streak angles from SAXS. (b) Profiles along the streaks of cold (blue) and HI laser pumped (orange) wires for those images from Fig.\ref{fig:SAXS} where the streaks can be clearly separated and are sufficiently intense. (c) Fitted change of the roughness parameter as function of probe delay.}
\label{fig:traces}
\end{figure*}

\begin{figure*}
\centering
\includegraphics[width=\linewidth]{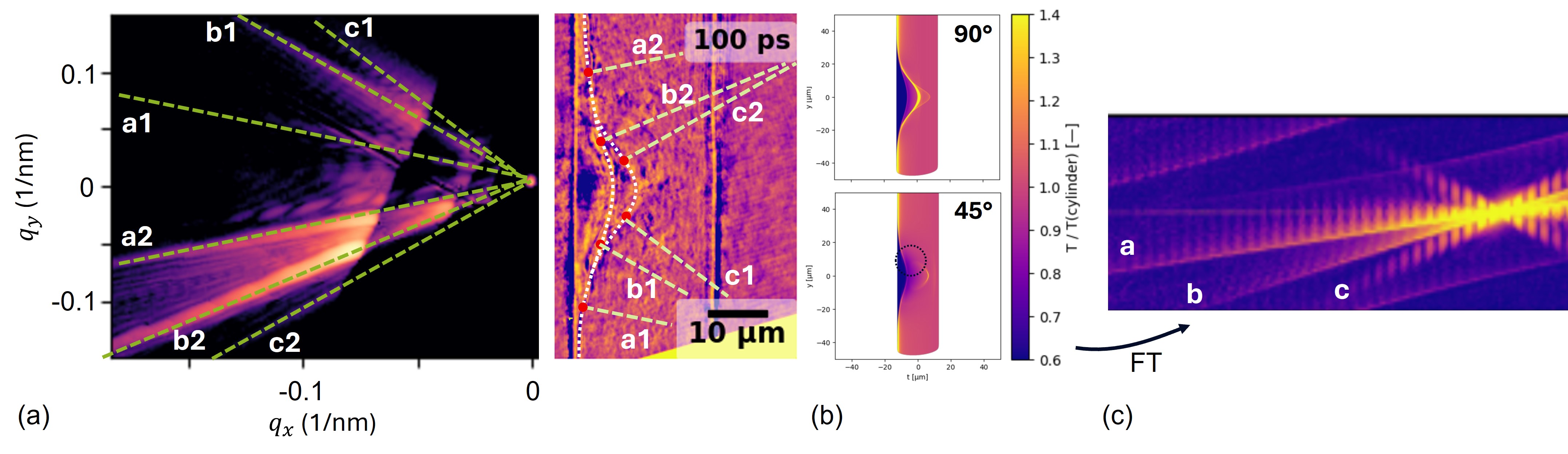}
\caption{Correlation between SAXS and XRM. 
(a) Overlay of SAXS streak orientations with the corresponding XRM-derived fronts.  
(b) Density model projection used to reproduce the observed transmission and scattering patterns by using a wide Gaussian (corresponding to streaks ai) and a narrower one (which produces two apparent fronts upon rotation, corresponding to streaks bi, ci). 
The combined analysis links specific SAXS features to the distinct compression fronts identified in XRM. (c) Fourier Transform of the density projection shown in (b) with the XFEL beam spot on target assumed to be transversely offset w.r.t. the laser axis (dashed lines in (b)). Note, that the high frequency oscillations are a numerical effect of the superposition of scattering from the perfectly aligned layers in the model.}
\label{fig:saxs-mxi}
\end{figure*}

The minimum and maximum streak angles as a function of delay are plotted in Fig.~\ref{fig:traces}(a). Streaks from day~1 are near the lower bound of those from day~2.
The absence of higher angle streaks suggests a small transverse offset between the XFEL probe and optical pump on day~1 since at larger distance to the HI laser the angle of the compression front seen in XRM are shallower, consistent with the lowest angle streaks (cp. Fig.~\ref{fig:saxs-mxi}. 
Using the XRM images as spatial reference, we can assign the observed streaks to specific regions of the target:  
D in Fig.~\ref{fig:SAXS}); larger-angle streaks (b, c) correspond to the more central, inward-curving fronts seen near the focal spot (B, C) (cp. Fig.~\ref{fig:saxs-mxi}).  
This spatial correspondence confirms that the SAXS signal originates from the same compression structures imaged by XRM, allowing quantitative inference of specific front sharpnesses. 

In the following we analyse the lower angle streaks, i.e. surfaces consistent with the transversely offset compression front (feature~D). 
We plot the profiles along streaks (a) for the pumped wires up to $300\unit{ps}$ probe delay in Fig.~\ref{fig:traces}(b), together with the respective profiles along the streak in the respective XFEL-only reference. 
We limit this analysis to data from day 1, since there the lower q data is less disturbed by overlap of neighboring streaks; and the signal strength of the streaks of interest was larger due to the better XFEL coverage. 
Note, however, that the same qualitative behaviour was observed on day~2, as well as for the streaks at larger angles. 
As can be seen, initially the slope of the pumped streak is slightly increased relative to the reference at $1\unit{ps}$ delay. At $6\unit{ps}$ the streak recovers and the slope is similar to the reference again. Later the slope sharply decreases between $45$ and $80\unit{ps}$ before it drops again.

To quantify the sharpness, we first benchmark against the cold (unpumped) reference wire scattering streaks, which follow $I(q)\propto\exp(-q^2\sigma_0^2)/q^3$, consistent with a smooth surface profile $\rho(x)\propto\mathrm{erf}(x/\sigma_0)$ and an average roughness $\sigma_0\approx10~\mathrm{nm}$ (see Methods). 
The pumped data were then fitted relative to this reference using
\begin{equation}
    R(q) = \frac{I_{\text{pumped}}(q)}{I_{\text{cold}}(q)} 
    \propto \exp\!\left[-q^2(\sigma^2-\sigma_0^2)\right],
\end{equation}
yielding the change in effective roughness $\Delta\Sigma\equiv\sign\sqrt{\sigma^2-\sigma_0^2}$ as a function of delay (Fig.~\ref{fig:traces}(c)). 
Initially ($t=1\unit{ps}$), $\Delta\Sigma>0$, consistent with slight surface expansion. 
At $6\unit{ps}$, $\Delta\Sigma$ approaches zero, indicating recovery. 
Subsequently $\Delta\Sigma$ becomes negative, reaching its minimum (i.e. sharpest surface) near $t=80\unit{ps}$. 
At this time, $\left|\Delta\Sigma\right|$ matches the cold-surface roughness, implying a fully developed, nanometre-sharp shock front. Beyond $100\unit{ps}$, the slope increases again, suggesting relaxation and decay of the shock --- and in agreement with the stagnation of the front seen in XRM.

The correspondence between model, XRM morphology, and SAXS-derived nanometre-scale roughness confirms that the observed compression front indeed represents a shock front.
Overall, the joint XRM–SAXS analysis bridges spatial scales from the micron level down to few nanometres, resolving the full evolution from macroscopic front geometry to microscopic interfacial sharpness. This combined diagnostic capability allows, for the first time, quantitative verification of nanometre-scale shock formation in a solid-density plasma driven by a Joule-class femtosecond laser pulse.

\subsection*{Shock formation}
The shock formation is not instantaneous, rather we find a considerable delay. 
At $1\unit{ps}$, $\Delta\sigma^2$ is negative, indicating expansion, while later it is first around zero and after $15\unit{ps}$ starting to become positive. 
We can estimate the breaking time assuming an isentropic flow of a fluid with local velocity $u(x,t)$, finite local sound speed $c(x,t)=\sqrt{\left(\partial p/\partial \rho\right)_s}$ where $s$ is the entropy of the fluid element, and with flux $j=\rho u$. 

\begin{figure}
\centering
\includegraphics[width=\linewidth]{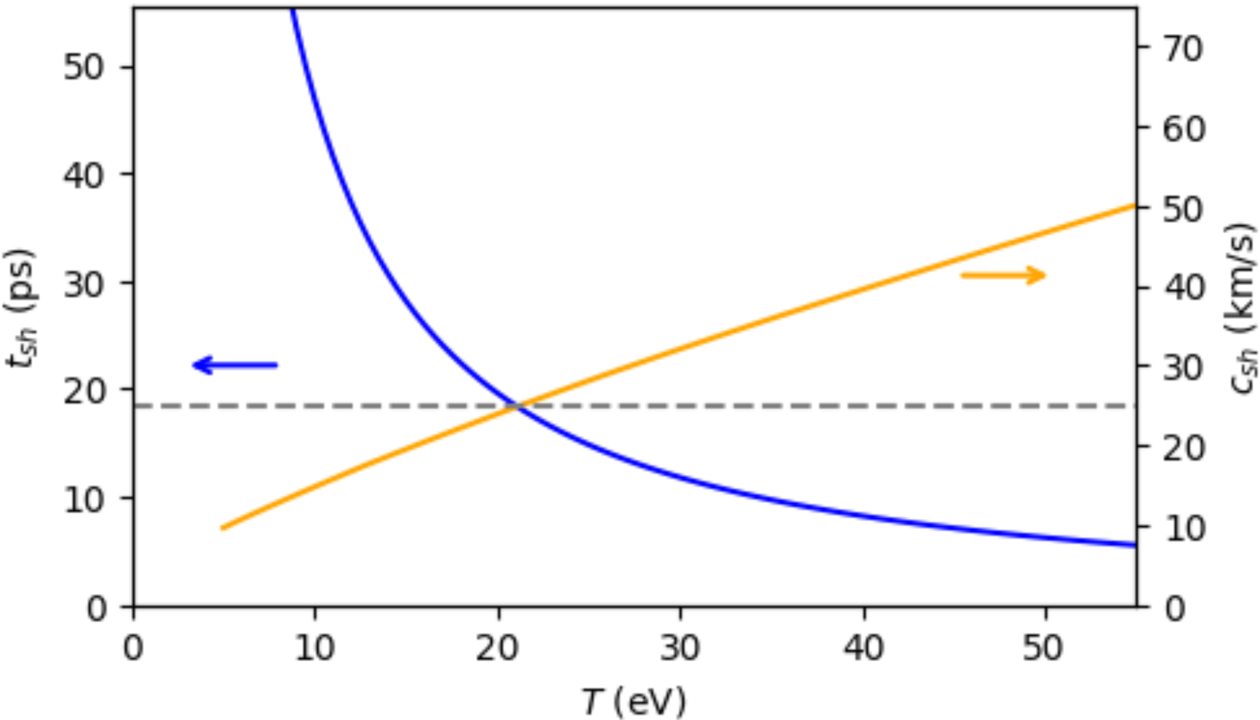}
\caption{$t_\mathrm{sh}$ (Eqn.~\eqref{eqn:t_sh_ideal}) and $c_{sh}$ (Eqn.~\eqref{eqn:c_sh}) for Copper (ideal gas approximation), simple wave, const. $Z(x,t)\cong \bar{Z}$ and Gaussian heat shape with width equal to the Drude skin depth.}
\label{fig:t_sh_c_sh}
\end{figure}

Neglecting heat flux, from the continuity equation and Euler momentum equations one can infer the shock formation time in the strong shock limit for an ideal gas ($r=\rho_{2}/\rho_1$=4, where $\rho_1$ is the upstream density, $\rho_2$ is the shock compressed one) (see Appendix)
\begin{align}
  t_{\mathrm{sh}} = 2e^{1/2}\sqrt{\gamma-1}\;\frac{\ell}{c_{\mathrm{sh}}}. 
  \label{eq:tsh_basic}
\end{align}
with the shock velocity
\begin{equation}
    c_{\mathrm{sh}}=\sqrt{\frac{r^2}{\left(r-1\right)} \frac{\left(\bar{Z} + 1\right) k_B T}{m_i}}. 
    \label{eqn:c_sh}
\end{equation}
where we assumed the ion charge to be approximately constant, $Z(x,t)\cong \bar Z$, and the heating profile to be Gaussian with peak temperature $T$ and width $\ell$.
To get an estimate for our experimental condition we can approximate $\ell$ with the Drude skin depth, the resulting shock formation time and shock velocity are plotted in Fig.~\ref{fig:t_sh_c_sh} as function of the temperature $T$ of the laser heated plasma. 
Specifically, from the imaging data we find that the shock front corresponding to the smallest angle streak has moved about $(2.5\pm 0.5)\unit{\mum}$ after $100\unit{ps}$, i.e. $c_{sh}\approx 25\unit{km/s}$ corresponding to a temperature of $k_B T_0 \approx (21.5\pm 2.5) \unit{eV}$. 
We then obtain the estimate $t_{sh}\approx (18\pm 3)\unit{ps}$ compared to $16\unit{ps}$ from a FLASH\cite{Fryxell2000} hydrodynamic simulation at that temperature. 
Considering the simplicity of the rough analytical estimate, the two are in good agreement with each other and with the SAXS results. 
There, a slight reduction of the slope can be seen after $15\unit{ps}$ for $q>0.08\unit{nm}^{-1}$, indicating already (partial) shock development. 
After $45\unit{ps}$ the streak is perpendicular to $1/q^3$ and at $80\unit{ps}$ it is almost perfectly following $1/q^3$, i.e. the shock is fully developed. 
Eventually at even later times the shock seems to decay as the slope increases again. 

\section*{Discussion and Conclusions}
This study demonstrates that combining XRM with SAXS yields a self-consistent, multi-scale view of laser-driven compression in solids—from the global morphology of hole-boring and compression fronts (microns) down to their nanometre-scale sharpness (few nanometres). 
XRM constrains the geometry and kinematics of the fronts, while SAXS isolates local interface sharpness through the asymptotic $I(q)\propto q^{-3}\exp(-q^2\sigma^2)$ fall-off and its temporal evolution. 
Taken together, these observables let us identify which XRM-visible interfaces generate the dominant SAXS streaks and convert streak slopes into an effective interfacial width $\sigma$, even when the real-space image cannot resolve the front.

Three central findings emerge.
First, the cold-wire reference shows an interfacial width $\sigma_0\approx10\,\mathrm{nm}$. During the pump–probe sequence, SAXS streaks associated with the laser-facing compression front initially broaden (negative $\Delta\sigma^2$ at $1\,\mathrm{ps}$, consistent with rarefaction), then steepen sharply between $30\,\mathrm{ps}$ and $80\,\mathrm{ps}$. By $t\approx80\,\mathrm{ps}$ the slope reaches the $q^{-3}$ limit characteristic of a discontinuous density jump, implying a fully developed shock. 
XRM confirms coincident front detachment, curvature, and compression, indicating that the apparent interface corresponds to a propagating shock rather than a smooth compression gradient.\\
Second, using the measured temperature profile and sound-speed gradient, the estimated breaking time $t_\mathrm{sh}\approx18\pm3\,\mathrm{ps}$ matches the observed SAXS evolution. 
This agreement supports an interpretation where the shock arises from nonlinear steepening of an initially smooth compression wave in the heated surface layer, rather than from instantaneous ablation-driven discontinuities. 
It is consistent with return current heating in a skin depth layer at the surface. \\
Identifying the front as a \emph{shock} is crucial because shocks mark the onset of irreversible thermodynamic processes such as entropy production, dissipative heating, and rapid pressure buildup. 
In a smooth compression wave, energy remains largely mechanical and reversible; the local temperature increases only adiabatically with density. 
A shock, by contrast, converts directed kinetic energy into internal energy. 
This entropy jump determines the downstream equation of state, the achievable pressure, and the efficiency of converting laser energy into compressive work.

In the present context, the existence of a true shock means that the ablation does not merely \emph{push} the material inward but \emph{dissipatively compresses} it. 
In the context of inertial confinement fusion energy, the ability to diagnose shock strength and sharpness directly enables validation of models of shock-induced heating and mix. 
Distinguishing a shock from a continuous compression wave validates the use of Rankine–Hugoniot relations, allowing for inference of temperature and density from front velocities and providing benchmarks for kinetic and hydrodynamic simulations.

\subsection*{Limitations}
This experiment closes the long-standing diagnostic gap between micron-scale imaging and nanometre-scale interface sensitivity. 
By combining MXI and SAXS, we can attribute specific SAXS features to real-space fronts and quantify their evolution. 
The $45^\circ$ pump–probe projection, limited angular sampling, and overlapping streaks constrain the precision of front localization. 
Limited sampling at higher $q$ values due to low signal and small detector active area limit the accuracy of the fits, which prohibits using more sophisticated fitting models. 
Hence, SAXS-derived $\sigma$ values are e.g. spatially averaged and underestimate the sharpest local gradients since the fits did not take into account partial shock formation or $\sigma$ variation across the relevant surface areas. 
Also, the used model of error-function density slopes is only a simple estimate and could be improved in the future, e.g. by using simulation-based profiles. 

\subsection*{Outlook}
The observation of a shock -- and its quantifiable formation dynamics -- has direct implications for inertial confinement fusion physics, warm dense matter studies, and high-energy-density material science. 
The MXI–SAXS platform provides e.g. a benchmark for validating resistive and ablative shock models including entropy production, a pathway to infer full Rankine–Hugoniot states from combined geometry, velocity, and sharpness data and a diagnostic foundation for exploring shock coalescence, radiative precursors, and turbulence onset in dense plasmas

Future extensions should integrate tomography or multi-view XRM to reconstruct three-dimensional shock geometry. 
Angle-resolved and energy-tuned SAXS could further distinguish electron-density and ionization-opacity contributions, enabling element-specific shock diagnostics.  

\section*{Acknowledgements}

Use of the Linac Coherent Light Source (LCLS), SLAC National Accelerator Laboratory, is supported by the U.S. Department of Energy, Office of Science, Office of Basic Energy Sciences under Contract No. DE-AC02-76SF00515. This research was supported by SC, Fusion Energy Science, FWP 100106: The LaserNetUS Initiative at Matter in Extreme Conditions, under Contract No. DE-AC02-76SF00515. We appreciate the support of HIBEF (www.hibef.eu). This research used resources of the National Energy Research Scientific Computing Center (NERSC), a Department of Energy User Facility using NERSC award instaplas-ERCAP~0036515. We also appreciate the support and use of resources of the HZDR high-performance data centre. The software used in this work was developed in part by the DOE NNSA- and DOE Office of Science-supported Flash Center for Computational Science at the University of Chicago and the University of Rochester. MF acknowledges US Department of Energy Office of Fusion Energy Sciences FWP100182. W.M.M. acknowledges support from the National Science Foundation Graduate Research Fellowship Program under Grant No. DGE-2146755. This work was funded by the DOE Office of Science, Fusion Energy Science under FWP100182.  C.G. acknowledges funding by the consortium DAPHNE4NFDI in association with the German National Research Data Infrastructure (NFDI) e.V. - project number 4602487. 
L.R. and C.G. acknowledge funding by the German Federal Ministry of Research, Technology and Space (BMFTR) Project No. 05K24PSA.

\section*{References}
\bibliographystyle{sn-mathphys-num}
\bibliography{export}

@article{Galtier2025,
   author = {Eric Galtier and Hae Ja Lee and Dimitri Khaghani and Nina Boiadjieva and Peregrine McGehee and Ariel Arnott and Brice Arnold and Meriame Berboucha and Eric Cunningham and Nick Czapla and Gilliss Dyer and Robert Ettelbrick and Philip Hart and Philip Heimann and Marc Welch and Mikako Makita and Arianna E. Gleason and Silvia Pandolfi and Anne Sakdinawat and Yanwei Liu and Michael J. Wojcik and Daniel Hodge and Richard Sandberg and Maria Pia Valdivia and Victorien Bouffetier and Gabriel Pérez-Callejo and Frank Seiboth and Bob Nagler},
   doi = {10.1038/s41598-025-91989-8},
   issn = {2045-2322},
   issue = {1},
   journal = {Scientific Reports},
   month = {3},
   pages = {7588},
   pmid = {40038475},
   publisher = {Nature Research},
   title = {X-ray microscopy and talbot imaging with the matter in extreme conditions X-ray imager at LCLS},
   volume = {15},
   url = {https://www.nature.com/articles/s41598-025-91989-8},
   year = {2025}
}

@book{Whitham1974,
   author = {Gerald Beresford Whitham},
   city = {New York},
   isbn = {0-471-94090-9},
   publisher = {Wiley-Interscience},
   title = {Linear and nonlinear waves},
   year = {1974}
}

@article{Ruhl2022,
   author = {Hartmut Ruhl and Georg Korn},
   month = {2},
   title = {A laser-driven mixed fuel nuclear fusion micro-reactor concept},
   url = {https://arxiv.org/pdf/2202.03170},
   year = {2022}
}

@article{Yang2024,
   author = {Long Yang and Martin Rehwald and Thomas Kluge and Alejandro Laso Garcia and Toma Toncian and Karl Zeil and Ulrich Schramm and Thomas E. Cowan and Lingen Huang},
   doi = {10.1063/5.0181321},
   issn = {2468-2047},
   issue = {4},
   journal = {Matter and Radiation at Extremes},
   month = {7},
   pages = {047204-},
   publisher = {Matter and Radiation at Extremes},
   title = {Dynamic convergent shock compression initiated by return current in high-intensity laser–solid interactions},
   volume = {9},
   url = {https://pubs.aip.org/mre/article/9/4/047204/3296069/Dynamic-convergent-shock-compression-initiated-by},
   year = {2024}
}

@article{Yang2025,
   author = {L. Yang and M. -L. Herbert and C. Bähtz and V. Bouffetier and E. Brambrink and T. Dornheim and N. Fefeu and T. Gawne and S. Göde and J. Hagemann and H. Höeppner and L. G. Huang and O. S. Humphries and T. Kluge and D. Kraus and J. Lütgert and J. -P. Naedler and M. Nakatsutsumi and A. Pelka and T. R. Preston and C. Qu and S. V. Rahul and R. Redmer and M. Rehwald and L. Randolph and J. J. Santos and M. Šmíd and U. Schramm and J. -P. Schwinkendorf and M. Vescovi and U. Zastrau and K. Zeil and A. Laso Garcia and T. Toncian and T. E. Cowan},
   month = {7},
   title = {Scaling of thin wire cylindrical compression after 100 fs Joule surface heating with material, diameter and laser energy},
   url = {http://arxiv.org/abs/2507.12109},
   year = {2025}
}

@article{Schoenwaelder2025,
   author = {C Schoenwaelder and A Marret and S Assenbaum and C B Curry and E Cunningham and G Dyer and S Funk and G D Glenn and S Goede and D Khaghani and M Rehwald and U Schramm and F Treffert and K Zeil and S H Glenzer and F Fiuza and M Gauthier},
   journal = {Nature Physics},
   title = {Time-resolved X-ray imaging of the current filamentation instability in solid density plasmas},
   year = {2025}
}

@article{Garcia2024,
   author = {Alejandro Laso Garcia and Long Yang and Victorien Bouffetier and Karen Appel and Carsten Baehtz and Johannes Hagemann and Hauke Höppner and Oliver Humphries and Thomas Kluge and Mikhail Mishchenko and Motoaki Nakatsutsumi and Alexander Pelka and Thomas R. Preston and Lisa Randolph and Ulf Zastrau and Thomas E. Cowan and Lingen Huang and Toma Toncian},
   doi = {10.1038/s41467-024-52232-6},
   issn = {2041-1723},
   issue = {1},
   journal = {Nature Communications},
   month = {9},
   pages = {7896},
   title = {Cylindrical compression of thin wires by irradiation with a Joule-class short-pulse laser},
   volume = {15},
   year = {2024}
}

@article{Kluge2023,
   author = {Thomas Kluge and Michael Bussmann and Eric Galtier and Siegfried Glenzer and Jörg Grenzer and Christian Gutt and Nicholas J. Hartley and Lingen Huang and Alejandro Laso Garcia and Hae Ja Lee and Emma E. McBride and Josefine Metzkes-Ng and Motoaki Nakatsutsumi and Inhyuk Nam and Alexander Pelka and Irene Prencipe and Lisa Randolph and Martin Rehwald and Christian Rödel and Melanie Rödel and Toma Toncian and Long Yang and Karl Zeil and Ulrich Schramm and Thomas E. Cowan},
   doi = {10.48550/arXiv.2302.03104},
   journal = {arXiv},
   month = {2},
   pages = {2302.03104},
   title = {Probing the dynamics of solid density micro-wire targets after ultra-intense laser irradiation using a free-electron laser},
   url = {http://arxiv.org/abs/2302.03104},
   year = {2023}
}

@article{Gaus2021,
   author = {Lennart Gaus and Lothar Bischoff and Michael Bussmann and Eric Cunningham and Chandra B. Curry and Juncheng E and Eric Galtier and Maxence Gauthier and Alejandro Laso García and Marco Garten and Siegfried Glenzer and Jörg Grenzer and Christian Gutt and Nicholas J. Hartley and Lingen Huang and Uwe Hübner and Dominik Kraus and Hae Ja Lee and Emma E. McBride and Josefine Metzkes-Ng and Bob Nagler and Motoaki Nakatsutsumi and Jan Nikl and Masato Ota and Alexander Pelka and Irene Prencipe and Lisa Randolph and Melanie Rödel and Youichi Sakawa and Hans-Peter Schlenvoigt and Michal Šmíd and Franziska Treffert and Katja Voigt and Karl Zeil and Thomas E. Cowan and Ulrich Schramm and Thomas Kluge},
   doi = {10.1103/PhysRevResearch.3.043194},
   issn = {2643-1564},
   issue = {4},
   journal = {Physical Review Research},
   month = {12},
   pages = {043194},
   publisher = {American Physical Society},
   title = {Probing ultrafast laser plasma processes inside solids with resonant small-angle x-ray scattering},
   volume = {3},
   url = {https://link.aps.org/doi/10.1103/PhysRevResearch.3.043194},
   year = {2021}
}

@article{Roth2001,
   author = {M. Roth and T. E. Cowan and M. H. Key and S. P. Hatchett and C. Brown and W. Fountain and J. Johnson and D. M. Pennington and R. A. Snavely and S. C. Wilks and K. Yasuike and H. Ruhl and F. Pegoraro and S. V. Bulanov and E. M. Campbell and M. D. Perry and H. Powell},
   doi = {10.1103/PhysRevLett.86.436},
   isbn = {0031-9007 (Print)\r0031-9007 (Linking)},
   issn = {00319007},
   issue = {3},
   journal = {Physical Review Letters},
   month = {1},
   pages = {436-439},
   pmid = {11177849},
   publisher = {APS},
   title = {Fast ignition by intense laser-accelerated proton beams},
   volume = {86},
   year = {2001}
}

@article{Kluge2016,
   author = {T. Kluge and M. Bussmann and H.-K. Chung and C. Gutt and L. G. Huang and M. Zacharias and U. Schramm and T. E. Cowan},
   doi = {10.1063/1.4942786},
   issn = {1070-664X},
   issue = {3},
   journal = {Physics of Plasmas},
   month = {3},
   pages = {033103},
   title = {Nanoscale femtosecond imaging of transient hot solid density plasmas with elemental and charge state sensitivity using resonant coherent diffraction},
   volume = {23},
   url = {https://pubs.aip.org/pop/article/23/3/033103/1018148/Nanoscale-femtosecond-imaging-of-transient-hot},
   year = {2016}
}

@article{Kluge2018,
   author = {Thomas Kluge and Melanie Rödel and Josefine Metzkes-Ng and Alexander Pelka and Alejandro Laso Garcia and Irene Prencipe and Martin Rehwald and Motoaki Nakatsutsumi and Emma E. McBride and Tommy Schönherr and Marco Garten and Nicholas J. Hartley and Malte Zacharias and Jörg Grenzer and Artur Erbe and Yordan M. Georgiev and Eric Galtier and Inhyuk Nam and Hae Ja Lee and Siegfried Glenzer and Michael Bussmann and Christian Gutt and Karl Zeil and Christian Rödel and Uwe Hübner and Ulrich Schramm and Thomas E. Cowan},
   doi = {10.1103/PhysRevX.8.031068},
   issn = {2160-3308},
   issue = {3},
   journal = {Physical Review X},
   month = {9},
   pages = {031068},
   title = {Observation of Ultrafast Solid-Density Plasma Dynamics Using Femtosecond X-Ray Pulses from a Free-Electron Laser},
   volume = {8},
   url = {http://arxiv.org/abs/1801.08404 https://arxiv.org/pdf/1801.08404.pdf https://link.aps.org/doi/10.1103/PhysRevX.8.031068},
   year = {2018}
}

@article{Fryxell2000,
   author = {B. Fryxell and K. Olson and P. Ricker and F. X. Timmes and M. Zingale and D. Q. Lamb and P. MacNeice and R. Rosner and J. W. Truran and H. Tufo},
   doi = {10.1086/317361},
   issn = {0067-0049},
   issue = {1},
   journal = {The Astrophysical Journal Supplement Series},
   month = {11},
   pages = {273-334},
   title = {FLASH: An Adaptive Mesh Hydrodynamics Code for Modeling Astrophysical Thermonuclear Flashes},
   volume = {131},
   url = {www.c3.lanl.gov/dquinlan/AMR]].html. https://iopscience.iop.org/article/10.1086/317361},
   year = {2000}
}

@article{Kluge2017,
   author = {T. Kluge and C. Rödel and M. Rödel and A. Pelka and E. E. McBride and L. B. Fletcher and M. Harmand and A. Krygier and A. Higginbotham and M. Bussmann and E. Galtier and E. Gamboa and A. L. Garcia and M. Garten and S. H. Glenzer and E. Granados and C. Gutt and H. J. Lee and B. Nagler and W. Schumaker and F. Tavella and M. Zacharias and U. Schramm and T. E. Cowan},
   doi = {10.1063/1.5008289},
   issn = {10897674},
   issue = {10},
   journal = {Physics of Plasmas},
   month = {10},
   pages = {102709},
   title = {Nanometer-scale characterization of laser-driven compression, shocks, and phase transitions, by x-ray scattering using free electron lasers},
   volume = {24},
   url = {http://aip.scitation.org/doi/10.1063/1.5008289},
   year = {2017}
}

@article{Tabak1994,
   author = {Max Tabak and James Hammer and Michael E. Glinsky and William L. Kruer and Scott C. Wilks and John Woodworth and E. Michael Campbell and Michael D. Perry and Rodney J. Mason},
   doi = {10.1063/1.870664},
   isbn = {1070-664X},
   issn = {1070664X},
   issue = {5},
   journal = {Physics of Plasmas},
   pages = {1626-1634},
   title = {Ignition and high gain with ultrapowerful lasers},
   volume = {1},
   url = {http://aip.scitation.org/doi/10.1063/1.870664},
   year = {1994}
}
  
\section*{Methods}

\subsection*{Fits}
All fits were performed by globally minimizing $\chi^2$, assuming a density following the cumulative distribution function $\Phi$ of a Gaussian (normal) distribution, $\rho\propto \Phi(x/\sigma)=0.5(1+\mathrm{erf}\left[x/(\sqrt{2}\sigma)\right]$. 
Then the intensity was fitted by $I=A \exp{\left(-q^2\sigma^2\right)}/q^3+const.$ ($\sigma \ll R$, where $R$ is the wire radius). 
\includegraphics[width=\linewidth]{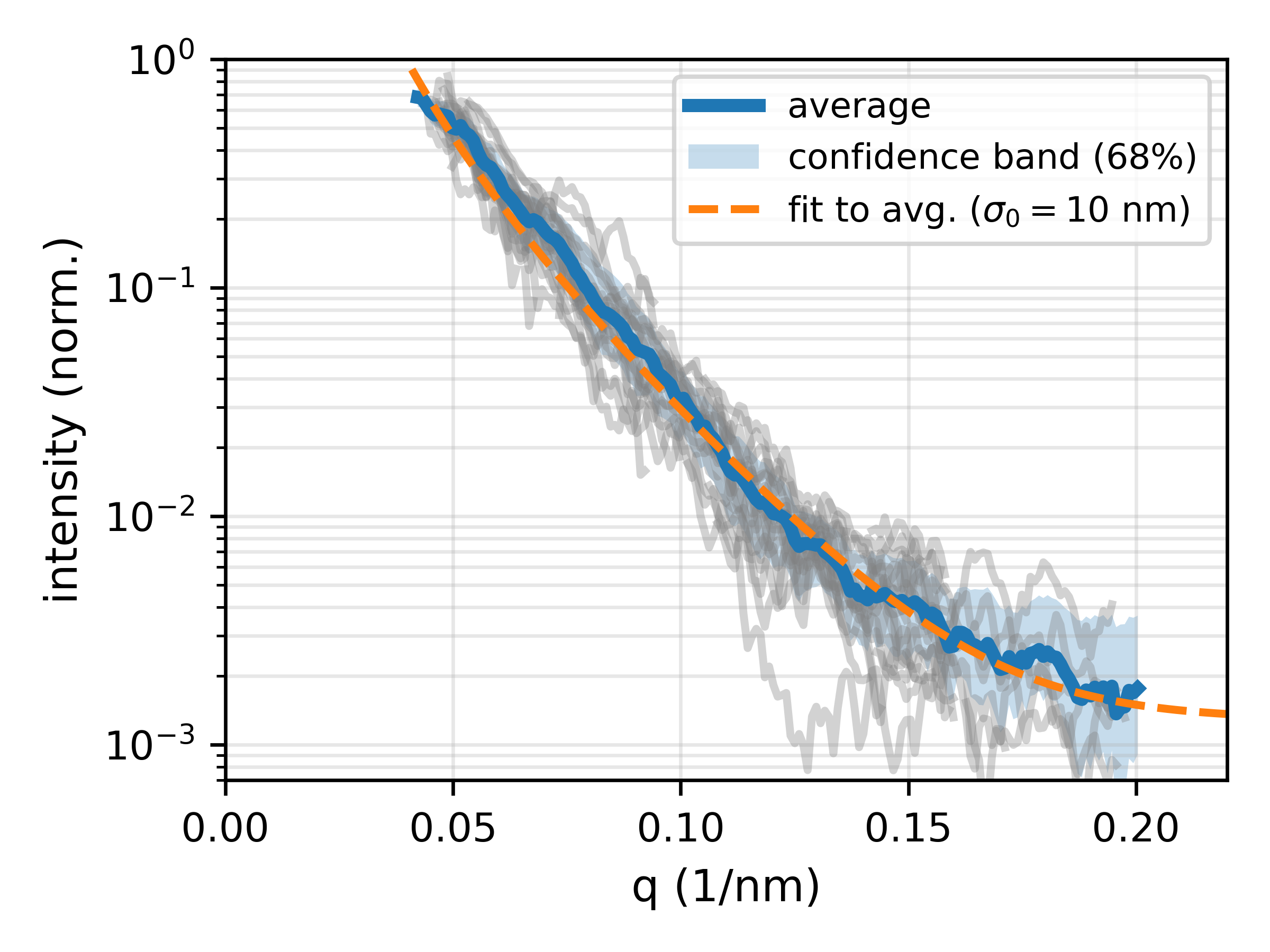}
\includegraphics[width=\linewidth]{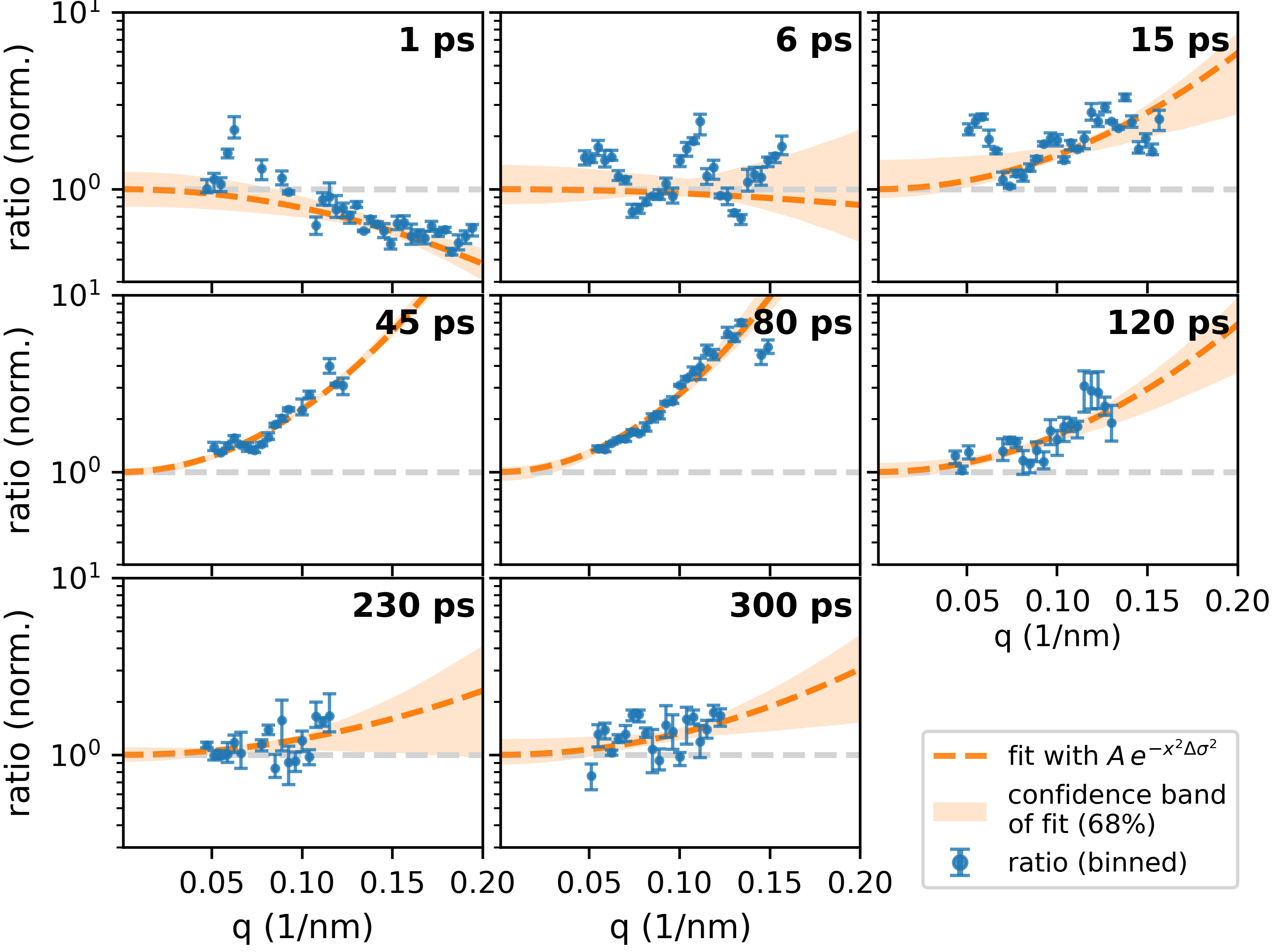}

\section*{Appendix}
For reference, we show the derivation of Eqn.~\eqref{eq:tsh_basic} following Whitham (1974)\cite{Whitham1974}. 
Neglecting heat flux, we start from the continuity equation 
\begin{align}
    \rho_t = -(\rho u)_x, \label{eq:cont1}
\end{align}
and the Euler momentum equation in its convective form
\begin{align}
    u_t = - u\,u_x - \frac{1}{\rho}\,p_x.\label{eq:mom1}
\end{align}
At this point the system consists of two coupled quasilinear PDEs in the variables $\rho$ and $u$. 
A standard way to analyze such systems is to search for curves in the space–time plane along which some combination $R$ of the variables $(\rho,u)$ is transported unchanged, i.e. $\dot{R}=0$. 
In other words, we are looking for curves $x(t)$ and combinations $R(\rho,u)$ such that the PDE system reduces to a transport equation described by the directional derivative along the slope $dx/dt=\lambda$ of the curve
\begin{align}
    (\partial_t+\lambda\,\partial_x)R(\rho,u)=0.\label{eqn:transport}
\end{align}
$R$ can be interpreted as information that is transferred with velocity $\lambda$ downstream, and the shock formation is equivalent to the information catastrophe when the mapping of particle initial positions $\zeta \mapsto x$ ceases to be invertible. 

We now rewrite Eqns.~\eqref{eq:cont1} and \eqref{eq:mom1} in directional derivative form. 
First, we expand the continuity equation, and add $\lambda\rho_x$ on both sides,
\begin{align}
 (\partial_t + \lambda\,\partial_x)\rho
 &= \big[-(u\,\rho_x + \rho\,u_x)\big] + \lambda \rho_x \nonumber\\
 &= (\lambda - u)\,\rho_x - \rho\,u_x. \label{eq:D_lambda_rho}
\end{align}
Likewise, we add $\lambda u_x$ on both sides of the momentum equation and replace $p_x=p_\rho \rho_x$ and $c^2(\rho) \equiv p_\rho$,
\begin{align}
 (\partial_t + \lambda\,\partial_x)u
 &= \big[-u\,u_x - \tfrac{c^2}{\rho}\rho_x\big] + \lambda u_x \nonumber\\
 &= (\lambda - u)\,u_x - \frac{c^2}{\rho}\,\rho_x. \label{eq:D_lambda_u}
\end{align}
We choose the Ansatz $R=u+f(\rho)$, then the total derivative
\begin{align}
  (\partial_t+\lambda\partial_x)R
  = (\partial_t+\lambda\partial_x)u + f'(\rho)(\partial_t+\lambda\partial_x)\rho
\end{align}
must vanish. 
Inserting the expressions above and requiring both coefficients of $\rho_x$ and $u_x$ to vanish in order to obtain Eqn.~\eqref{eqn:transport} yields the system
\begin{align}
  -\frac{c^2}{\rho}+f'(\rho)(\lambda-u)&=0,\\
  (\lambda-u)-f'(\rho)\rho&=0.\label{eqn:R_eqn}
\end{align}
Eliminating $f'(\rho)$ between the two conditions shows that the two possible characteristic slopes are
\begin{align}
  \lambda_\pm = u \pm c. \label{eq:char-slopes}
\end{align}
For either choice, relation \eqref{eqn:R_eqn} can be integrated for
\begin{align}
   f(\rho) = \int \frac{\pm c(\rho)}{\rho}\,d\rho+const.
\end{align}
and without loss of generality we set 
\begin{align}
   R_\pm(\rho,u) = u \pm \int^\rho \frac{c(\rho')}{\rho'}\,d\rho'.
\end{align}

For a polyropic EOS with $p\propto \rho^\gamma$, $c(\rho)=\sqrt{\gamma p/\rho}$ the integral can be evaluated explicitly, which yields 
\begin{align}
  R_\pm = u \pm \frac{2}{\gamma-1}c.
\end{align}

In a forward-running simple wave we set $\lambda\equiv\lambda_+$, $R_-$ is uniform and only $R\equiv R_+$ varies. 
Then characteristics from different starting positions $\zeta$ move as
\begin{equation}
    x(t,\zeta) \approx \zeta + \lambda(R)\,t
\end{equation}
with the frozen velocity map $\lambda(R)\equiv \lambda(R(\zeta,0))$ since $R$ is constant. 
A shock forms when
\begin{equation}
    \frac{\partial x}{\partial \zeta} = 0 \;\;\Longrightarrow\;\; t_{sh}(\zeta) = -\frac{1}{\lambda'_0(R) R_\zeta},
    \label{eq:t_sh_general}
\end{equation}
i.e. the shock has fully formed at $t_{sh} = -1/\mathrm{min}_\zeta \big(\lambda'_0(R) R_\zeta\big)$. 
The blow-up of the gradient of the density $\rho(R)$ can be seen explicitly by setting $r=R_x$. 
Differentiating the scalar transport equation \eqref{eqn:R_eqn}
with respect to $x$ yields
\begin{align}
   r_t + \lambda(R) r_x + \lambda'(R)\,r^2 = 0.
\end{align}
Along a characteristic $dx/dt=\lambda(R)$, $R$ is constant and the equation for $r$ reduces to
\begin{align}
   \frac{dr}{dt} + \lambda'(R)\,r^2 = 0,
\end{align}
with solution
\begin{align}
   \frac{1}{r(t)} = \frac{1}{r(0)} + \lambda'(R)\,t.
\end{align}
At $t=t_{sh}$ the right hand side vanishes
and hence $r(t) \to \infty$. 
As $R$ develops a jump, the density $\rho(R)$ does the same, indicating the usual shock front condition. 

We now continue to write the shock appearance time explicitly for the ideal gas approximation. 
Using the explicit dependence 
\begin{align}
     \lambda = u+c = \frac{\gamma+1}{4} R + \frac{3-\gamma}{4} R_-,
\end{align}
the denominator in \eqref{eq:t_sh_general} becomes
\begin{align}
    \lambda'_0(R)\,R_\zeta = \frac{\gamma + 1}{4}\left(u_0'(\zeta)+\frac{2}{\gamma-1} c_0'(\zeta)\right).
\end{align}
For impulsive surface heating, $\tau_{\mathrm{laser}}\ll t_{sh}$, the initial flow gradient is negligible so that $u_0'(\zeta)\approx 0$. 
Thus the breaking time reduces to
\begin{equation}
    t_{sh} = -\,\frac{2(\gamma-1)}{\gamma+1}\;\frac{1}{\min_x\big[c_0'(x)\big]}.
    \label{eqn:t_sh}
\end{equation}

With 
\begin{equation}
    c_0(x) =\sqrt{\gamma \frac{\left(Z(T,n_e) +1\right) k_B T(x)}{m_i}}
\end{equation}
and the approximation of a Gaussian temperature shape with width equal $\ell$
\begin{equation}
    T(x) = T_0\cdot e^{-\frac{(x-x_0)^2}{2\ell^2}}
\end{equation}
and approximating $Z=\bar{Z}$ from Saha-equilibrium as constant across the compression wave, it is
\begin{align}
  c_0'(x)=-\frac{x-x_0}{2\ell^2}\;c_0(x).
\end{align}
The most negative slope occurs at $|x-x_0|=\sqrt{2}\,\ell$, giving
\begin{align}
  \min_x c_0'(x)=-\frac{1}{\sqrt{2}\,\ell}\,c_{0}(0)\,e^{-1/2}.
\end{align}
Inserting this into Eqn.~\eqref{eqn:t_sh} then yields the shock formation time
\begin{align}
  t_{\mathrm{sh}}
  =\frac{2(\gamma-1)}{\gamma+1}\;\sqrt{\frac{2 \exp(1) m_i}{ \gamma (\bar{Z}+1) k_B T_0}}
  \ell
  \label{eqn:t_sh_ideal}
\end{align}
which in the strong shock limit can be simply written as
\begin{align}
  t_{\mathrm{sh}} = 2e^{1/2}\sqrt{\gamma-1}\;\frac{\ell}{c_{\mathrm{sh}}}. 
  \label{eq:tsh_final}
\end{align}
Here $c_{\mathrm{sh}}$ is the shock velocity which can be obtained from the Rankine-Hugeniot condition. 
The compression factor and pressure in the shocked region (index 2) in the shock co-moving frame are $r=\rho_{2}/\rho_1=\left(\gamma+1\right)/\left(\gamma-1\right)$, $p_{2}=2/\left(\gamma+1\right)\rho_1 c_{\mathrm{sh}}^2$ and with the ideal gas EOS 
it is straight forward to obtain the shock velocity in the laboratory frame
\begin{equation}
c_{\mathrm{sh}}=\sqrt{\frac{r^2}{\gamma\left(r-1\right)}} c_0. 
\label{eqn:c_sh2}
\end{equation}

\begin{figure}
\centering
\includegraphics[width=\linewidth]{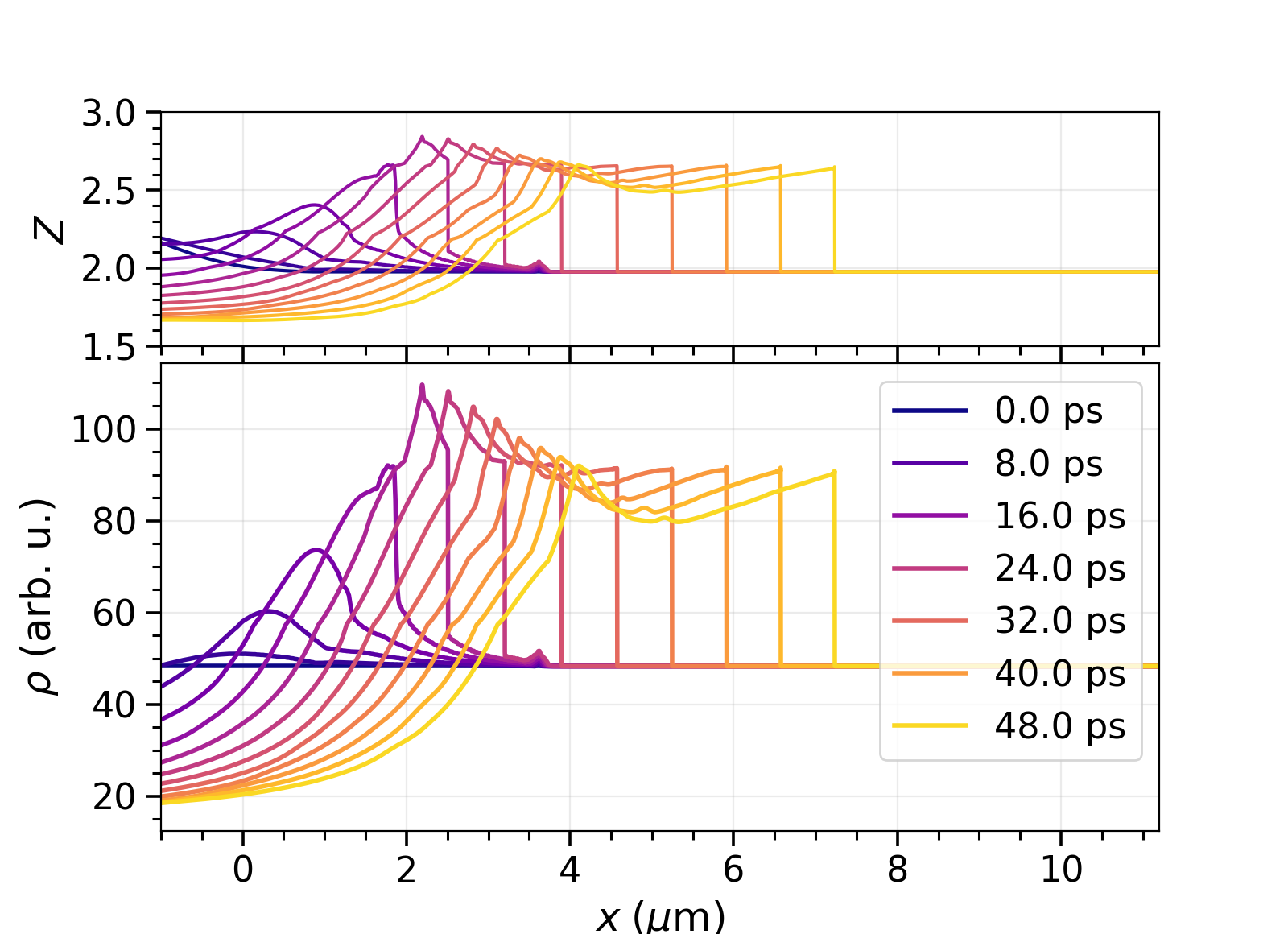}
\caption{$\rho(t)$ from hydrodynamic simulation (FLASH\cite{Fryxell2000}). Profiles are drawn every $4\unit{ps}$, only every other line colour is labeled.}
\label{fig:rho_t_FLASH}
\end{figure}

To get an estimate for our experimental condition we can approximate $\ell$ with the Drude skin depth $$\ell_s\approx\Im{\left[\frac{\omega}{c}\sqrt{1-\frac{\omega_p^2}{\omega\left(\omega+i\nu_0\right)}}\right]}^{-1}$$ (where $\omega_p=\left(n_e e^2/m_e \varepsilon_0\right)^{1/2}$ is the electron plasma frequency, $n_e = \bar{Z} n_i$ is the free electron density, $\nu_0 = \varepsilon_0\omega_p^2~\eta_0$ is the collision frequency, and $\eta_0=4\sqrt{2\pi}/3\cdot e^{1/2}m_e^{1/2}/4\pi\varepsilon_0\cdot Z \ln{\mathrm{\Lambda}}/T_0^{3/2}$ is the Spitzer resistivity) with respect to the heating return current pulse with effective frequency $\omega=2\pi/\tau$
\cite{Garcia2024,Yang2024,Yang2025}. 
The resulting shock formation time and shock velocity are plotted in Fig.~\ref{fig:t_sh_c_sh} as function of $T_0$.
Specifically, from the imaging data we find that the shock front corresponding to the smallest angle streak has moved about $(2.5\pm 0.5)\unit{\mum}$ after $100\unit{ps}$, i.e. $c_{sh}\approx 25\unit{km/s}$ corresponding to a temperature of $k_B T_0 \approx (21.5\pm 2.5) \unit{eV}$. 
Then the average ionisation is approximately $2.5-3$ from Saha equilibrium with 1 electron in the conduction band at room temperature, and the skin depth is approximately $(215\pm 10)\unit{nm}$. 
We then finally obtain the estimate $t_{sh}\approx (18\pm 3)\unit{ps}$. \\
Figure~\ref{fig:rho_t_FLASH} shows the simulated 1 dimensional temporal evolution of the density and ion charge state for $T_0=21.5\unit{eV}$, Gaussian profile with width equal to the skin depth assuming $45\unit{fs}$ pulse duration using the hydrodynamic simulation code FLASH\cite{Fryxell2000}. 
Shock formation occurs at approx. $16\unit{ps}$ which is in reasonable agreement with the analytic value. 

\end{document}